**Some formulae for evaluation of the polariton and plasmon frequncies, in materials.**

Voicu Dolocan[1], Andrei Dolocan[2, *], and Voicu Octavian Dolocan[3]

1. *Faculty of Physics, University of Bucharest, Bucharest, Romania*
2. *National Institute for Laser, Plasma and Radiation Physics, Bucharest, Romania*
3. *Aix-Marseille University &IM2NP, Avenue Escadrille Normandie Niemen, 13397 Marseille cedex 20, France*

**Abstract.** We present new relations for calculation of the phonon polaritons, plasmons and plasmon polaritons frquencies, which differ from the standard relations. So, we study the frequency of phonon polaritons in a two dimensional lattice with applications to $CuO_2$ lattices. Next, we evaluate plasmon frequencies in graphene and in $Cu2D$ and $Cu3D$ metals. Further, we evaluate the frequency of surface plasmon polaritons at the $Cu2D$/air and $Cu3D$/air interfaces.The obtained results are in a good agreement with experimental data. Also, we present a correspondence between the potential vector and the displacement of an oscillator. It is well known that the commutation relations for the operators in second quantization originate from the simple harmonic oscillators that are used to quantize the electromagnetic field. Finally, we present a comparison between photons, phonons and plasmons.

## I. INTRODUCTION

Polaritons and plasmons are both quantum-mechanical quasiparticles which are used to describe interactions in a solid. Polaritons are quasiparticles resulting from strong coupling of electromagnetic waves with an elecric or magnetic dipole carying excitation. Polariton is boson (not to be confused with the polaron, which is a fermion). At the point where the two dispersion relationships of light and excitation are crossing each other they have the same energy and therefore coupling occurs. In a system of two coupling oscillators with two natural frequencies, as the coupling strength increases, the lower frequency decreases and the higher increases. The effect can be viewed as a repulsion between the frequencies.

A plasmon is a quasiparticle resulting from the quantization of plasma oscillations of the free electron gas density with respect to the fixed positive ions. Plasmon is a boson. The free electrons in a metal may be considered an electron plasma, and the optical properties of metals can be described with plasmons. Light of frequency above the plasma frequency is transmitted, because the electrons cannot respond fast enough to screen it. Light of frequency below the plasma frequency is reflected, because the electrons in the metal screen the electric field of the light. Most metals and semiconductors are reflective in the visible range because their plasmon frequency is in the ultraviolet. Some metals, such as copper and gold, have electronic interband transitions in the visible range, whereby specific frequencies (color) are absorbed. Thus, these metals have a distinct color.

There are different types of interaction: phonon-polariton interaction coupling of infrared photons with optical phonons), surface plasmon-polariton interaction ( coupling of surface plasmons with light), exciton-polariton interaction ( coupling of photons of visible light with an exciton). The phonon-polariton can be found in ionic crystals (mostly insulators) such as NaCl and they only occur in materials with 2 or more atoms per unit cell, because otherwise there are no optical phonons.

The quantum theory of the interaction of radiation with optical phonons was first presented by Fano[1] and Hopfield[2]. Also, a quantum mechanical treatment of polaritons and plasmons is presented in Ref. [3]. The problem of polaritons and plasmons is studied in many papers[4–10]. In this paper, we continue the studies presented in Ref. [3]. First, we demand and find a justification to apply our model to the electromagnetic field. Then, we present some application to polaritons and plasmons.

## II. MODEL AND FORMULATION

The Hamiltonian density of a system of two bodies interacting via a boson field is

$$H_d = \frac{1}{2\rho}\Pi_\mu \Pi_\mu + \frac{D_l R_l}{2}\frac{\partial u_\mu}{\partial z_\mu}\frac{\partial u_l}{\partial z_l} + \frac{D_t R_l}{2}\frac{\partial u_\mu}{\partial z_l}\frac{\partial u_\mu}{\partial z_{ol}} + \frac{D_l s_l}{2}\frac{\partial u_\mu}{\partial z_\mu}\frac{\partial u_l}{\partial z_l} + \frac{D_t s_l}{2}\frac{\partial u_\mu}{\partial z_l}\frac{\partial u_\mu}{\partial z_l} \qquad (1)$$

We are to sum over repeated incices. The coordinate axes $z_\mu$ are assumed to be orthogonal. The term in $D_l$ is the square of the trace of the strain tensor; the term in $D_t$ is the sum of the squares of the tensor components. $R_l$ is the distance between the two bodies, $s_l$ is the relative displacement of the two bodies orieneted along the axis connecting them,

$$\rho = \rho_0 + (D_l \delta\mu l + D_t)\frac{R_l}{c^2} = \rho_o + \frac{DR_l}{c^2} \qquad (2)$$

The last term from the right hand side is the density of the coupling field, $\Pi_\mu$ are the components of the momentum density, $u_\mu(z)$ are the displacements of the coupling field at the position $z$

$$u_\mu(z) = \frac{1}{\sqrt{NR}}\sum \left(\frac{\hbar}{2\rho \omega_{k\mu}}\right)^{1/2}(a_{k\mu}e^{ikz} + a^p_{k\mu}e^{-ikz}) \qquad (3)$$

where the summation is performed over $k$. Further,

$$s_l(z) = \frac{1}{N}\sum Q_q e^{iq(z-z_n)}$$

$$Q_q = \left(\frac{\hbar}{2m\Omega_q}\right)^{1/2}(b_q + b^p_{-q})$$

where the summation is performed over $q$ and $n$. $\omega_{k\mu}$ is the classical oscillation frequency

$$\omega_{k\mu} = \left(\frac{D_l \delta_{\mu l} + D_t}{\rho}\right)^{1/2} k = \left(\frac{D}{\rho}\right)^{1/2} k \qquad (4)$$

where $l$ denotes longitudinal boson and other two choices of $\mu$ denote transverse bosons. The Hamiltonian of interaction is

$$H_I = \frac{D_l}{2}\int \sum s_l \frac{\partial u_\mu}{\partial z_\mu}\frac{\partial u_l}{\partial z_l}dz + \frac{D_t}{2}\int \sum s_l \frac{\partial u_\mu}{\partial z_l}\frac{\partial u_\mu}{\partial z_l}dz \qquad (5)$$

where we will sum over the neighbours, $l$. The total Hamiltonian of the interacting bodies and the connecting field is

$$H = H_o + H_I$$

$$H_o = H_{o1} + H_{o2} = \sum \hbar\Omega_q b^p_q b_q + \sum \hbar\omega_k a^p_k a_k \qquad (6)$$

where the summation is performed over $q$ and $k$, respectively. $\Omega_q, \omega_\kappa$ are just the classical oscillations frequencies, $a_k^p$, $a_k$ and $b_q^p$, $b_q$ are boson creation and annihilation operators for the connecting filed and the interacting bodies, respectively, N is the number of the bodies which have the same mass $m$ and in the equillibrium positions, lie at a distance $R$ from one another, $\rho_o R = m_o$ is the mass associated with the interacting field if this is a massive field, $c$ is the light velocity and $D$ is a coupling constant. It is assumed that in the approximation of the nearest neighbours, D does not depend on $l$. May be written

$$H_I = \frac{D}{2} \sum \left(\frac{\hbar}{2m\Omega_q}\right)^{1/2} (b_q + b_{-q}^p) e^{-iqz_n} \frac{1}{NR} \frac{\hbar k k'}{2\rho(\omega_k \omega_{k'})^{1/2}} (a_k + a_{-k}^p)(a_{-k'} + a_{k'}^p) \int e^{i(k-k'+q)z} dz \qquad (7)$$

where the summation is performed over *n, k, k', q.* Consider the integral over *z*

$$\int e^{i(k-k'+q)z} dz = NR \delta(k - k' + q)$$

and therefore $k' = k + q$. We write

$$(a_k + a_{-k}^p)(a_{-(k+q)} + a_{k+q}^p) = a_k a_{-(k+q)} + a_k a_{k+q}^p + a_{-k}^p a_{-(k+q)} + a_{-k}^p a_{k+q}^p \qquad (9)$$

First, for the sake of convenience, we assume the following situation: initially, there is a boson in *k* and none in *k + q*; then there is one in *k + q* but not in *k*. Obviously, the terms with two annihilators or two creators do not contribute, so that we omit them. Further, in the case of the weak coupling, we assume

$$n_q = \sum a_{k+q}^p a_k = \sqrt{N_o}(a_{-q} + a_q^p)$$

so that

$$\sum (a_k + a_{-k}^p)(a_{-(k+q)} + a_{k+q}^p) = 2\sqrt{N_o}(a_{-q} + a_q^p)$$

where the summation is performed over *k*. $N_o = \langle\phi(0)|a_o^+ a_o|\phi(0)\rangle$ is the average number of partices in the zero momentum state and the number of the particles in the system is the expectattion value of $N_o + \sum' a_{k+q}^p a_k$ where the sign prime shows that the terms for *k* = 0 and *k + q* = 0 are omitted. This consideration refers to phonons. The Hamiltonian of interaction may be written

$$H_I = \sum F_{k,q}(b_q a_{-q} + b_q a_q^p + b_{-q}^p a_{-q} + b_{-q}^p a_q^p) \qquad (10)$$

where the summation is performed over *k* and *q*, and

$$F_{k,q} = \frac{D}{2}\left(\frac{N_o}{N}\right)^{1/2} \frac{\hbar k(k+q)}{\rho(\omega_k \omega_{k-q})^{1/2}} \left(\frac{\hbar}{2m\Omega_q}\right)^{1/2} \left(\frac{1}{N^{1/2}} \sum e^{-iqz_n}\right) \qquad (10a)$$

where the summation is performed over *n*.
Further, we choice how the interaction via electromagnetic field may be considered as an elastic coupling through flux lines. The electromagnetic field is characterized by electric field *E* and magnetic filed *H*. Because *H* has zero divergence, we may always take it to be the rotation of a vector field, i. e. the curl of the vector potential

$$H = \nabla \times A$$

By using Maxwell equations, *E* can be written in the formulation

$$E = \frac{-1}{c} \frac{\partial A}{\partial t} - gradV$$

where $V$ is the electric scalar potential, which may be choosen to be a constant. $A$ can be decomposed into plane waves as

$$A(r) = \sum \sqrt{\frac{\hbar 2\pi c^2}{\omega_k}} \frac{1}{\sqrt{V}} e_{k,j}(a_{k,j} + a^p_{-k,j}) e^{ikz} \tag{11}$$

As $A$ is a vector we have introduced an additional suffix $j$ for the polarization vector $e$ in addition to the wave vector $k$ when decomposing into plane waves, where $j$ can take the values 1 or 2. We have described the polarization vector by $e_{k,j}$. The Coulomb gauge

$divA = 0$

guarantees that the vector potential $A$ contains only transverse waves. Because of the transversality of $A$

$$e_{k,j} k = 0$$

We may further assume that the directions of polarization are at right angles to one another, i. e. that the relationship

$$e_{k,1} e_{k,2} = 0$$

applies. Next, we try find a correspondence between the vector potential and the displacement $u$ of an oscillator ( For more detais see Appendix). The classical expression for the energy of the electromagnetic field is

$$U = \frac{1}{8\pi} \int (E^2 + H^2) dz = \frac{1}{8\pi} \frac{1}{c^2} \int \left(\frac{\partial A}{\partial t}\right)^2 dz + \frac{1}{8\pi} \int (\nabla \times A)^2 dz = \frac{1}{4\pi} \frac{1}{c^2} \int \left(\frac{\partial A}{\partial t}\right)^2 dz$$

The classical expresion of the mechanical energy of the oscillator is

$$U = \frac{1}{2} \int \rho \left(\frac{\partial u}{\partial t}\right)^2 dz + T = \int \rho \left(\frac{\partial u}{\partial t}\right)^2 dz$$

where $\partial u/\partial t$ is the velocity and T is the potential energy. Comparing the two energy expessions, one obtains the following correspondence

$$u = \left(\frac{1}{4\pi c^2 \rho}\right)^{1/2} A = \left(\frac{\hbar}{2\rho\omega}\right)^{1/2} \frac{1}{\sqrt{V}} \sum e_{k,j}(a_{k,j} + a^p_{-k,j}) e^{ikr}$$

where the summation is performed over $k$ and $j$. Because the electromagneic wave is a transverse wave, we retain from the expresson (5) only the last term, which now we write it as

$$H_I = \frac{1}{2} D_t \int \sum s_l (grad\, \boldsymbol{u})^2 = \frac{1}{2} D_t \int \sum s_l [(grad\, u_x)^2 + (grad\, u_y)^2 + (grad\, u_z)^2] d\boldsymbol{z} = ¿$$
$$\frac{1}{2} D_t \frac{1}{4\pi c^2 \rho} \int \sum s_l [(grad\, A_x)^2 + (grad\, A_y)^2 + (grad\, A_z)^2] d\boldsymbol{z} \quad (12)$$

from which one obtains espression (7) if we assume that the wave propagates in **z** direction and $k_z^2 \sim k^2/3 \sim (1/3)kk'$. The summation is performed over $l$. On the basis of these consierations, we assume that the Hamiltonian (10) may be applied to study the interaction between phonons and photons.

### III. DIELECTRIC FUNCTION AND POLARITONS

Let us consider the interaction of the photons with optical phonons. This interaction is particularly important when the frequencies and wave vectors of the phonon and photon fields coincide. Near the crossover of the dispersion relations even weak coupling of two fields can have drastic effects. Such effects in isotropic or cubic crystals occur only for transverse optical phonons, as only these couple with an electromagnetic field, which is always transverse in isotropic media. Therefore, we assume that our system is formed from phonons and photons with **k = q**. The Hamiltonian of the system is

$$H = \sum \{\hbar \Omega_q b_q^p b_q + \hbar \omega_k a_k^p a_k + F_{k,q}(b_q a_{-q} + b_q a_q^p + b_{-q}^p a_{-q} + b_{-q}^p a_q^p)\}$$

It may be diagonalized and one obtains

$$\omega^4 - \omega^2(\omega_o^2 + c^2 q^2) + \omega_o^2 c^2 q^2 - \frac{4\omega_o cq |F|^2}{\hbar^2} = 0$$

where $\omega_o = \omega_k$ is the optical phonon frequency, which is the longitudinal optical frequency, $\omega_L$ This relation may be written as

$$\omega^4 - \omega^2(\omega_L^2 + \frac{c^2 q^2}{\varepsilon_\infty}) + \frac{c^2 q^2}{\varepsilon_\infty}[\omega_L^2 - \frac{4\omega_L |F|^2 \sqrt{\varepsilon_\infty}}{\hbar^2 cq}] = 0 \quad (13)$$

To include the electronic polarizability we will replace $c^2$ by $c^2/\varepsilon_\infty$ where $\varepsilon_\infty$ is the square of the optical refractive index. By comparing this equation with standard equation

$$\omega^4 - \omega^2(\omega_L^2 + \frac{c^2 q^2}{\varepsilon_\infty}) + \frac{c^2 q^2 \omega_T^2}{\varepsilon_\infty} = 0 \quad (13a)$$

we identify

$$\omega_T^2 = \omega_L^2 - \frac{4\omega_L |F|^2 \sqrt{\varepsilon_\infty}}{\hbar^2 cq} \quad (14)$$

Thie relation may be compared with standard relation which gives the splitting between yhe frequencies of the longitudinal optical (LO) and transverse optical (TO) phonons in diatomic crystals.[11]

$$\omega_L^2 - \omega_T^2 = \frac{Ne^2 \varepsilon_s}{M_r \varepsilon_\infty}$$

where e* is an effective electron charge per ion [12] and is defined by

$$e^{*2} = M_r \Omega_o \omega_o^2 \varepsilon_o \left(\frac{1}{\varepsilon_\infty} - \frac{1}{\varepsilon_s}\right)$$

N is the number of unit cells per unit volume, $\varepsilon_s$ is the low frequency (static) dielectric constant, $M_r$ is the reduced mass of the two atoms, $\Omega_o$ is the unit cell volume, $\omega_o^2 = 2\beta/M_r$, $\beta$ is the force constant of the material. The difference betwen longitudinal and transverse frequencies at $k = 0$, arises from the ionicity of the crystal. For group !V semiconductors, there is of course, no splitting at $k = 0$. The above expressions are defined on the basis of the concept of macroscopic electric polarization. From the above expressions we see that the difference $\omega_L^2 - \omega_T^2$ depends on the wave vector $q$. When the phonon field is the interacting field between two photons, we replace in Eq. (10) $\rho = \rho_o$, $\omega_\kappa = \omega_T$, the tramsverse optical frequency, $\Omega_q = cq/\sqrt{\varepsilon_\infty}$, $\rho_o R = M_r$ (the reduced nass of the two interacting bodies), $m = \hbar q \sqrt{\varepsilon_\infty}/c$, one obtains

$$|F|^2 = \frac{D^2 \hbar^2 R^2 q^2}{2 \omega_k^2 M_r^2} \frac{N_o}{N} \left|\frac{1}{\sqrt{N}} \sum_n e^{i q z_n}\right|^2 \qquad (15)$$

When the photon field is the interacting field between two phonons we replace $\rho = \hbar q \sqrt{\varepsilon_\infty}/Rc$, $m = M_r$, $\omega_k = cq$, $\Omega_q = \omega_T$, and

$$|F|^2 = \frac{\hbar D^2 R^2}{4 M_r \omega_T} \frac{N_o}{N} \left|\frac{1}{\sqrt{N}} \sum_n e^{i q z_n}\right|^2 \qquad (16)$$

Now, we assume that both processes of interaction, between two photons via phonons and vice-versa, occur simultaneous. In this case for example $|F|^2$ is a geometrical average of $|F|^2$ (14) and $|F|^2$ (15) and one obtains

$$\omega_L^2 - \omega_T^2 = \frac{\sqrt{2} D^2 \omega_L R^2 \sqrt{\varepsilon_\infty}}{c \sqrt{\hbar} M_r^{3/2} \omega_T^{3/2}} \left|\frac{1}{\sqrt{N}} \sum_n e^{i q z_n}\right|^2 \qquad (17)$$

which, for $qR \ll 1$ do not depend on $q$. In this case, for a cubic crystal the modulus square in this expression equals 12 for NaCl structure, and equals 16 for CsCl structure. From te above relation may be determined the coupling constant D. The data obtained for the ionic crystals and III-V semiconductor compounds are in a good agreement with the experimentasl data for bulk modulus[3,13]. By using Eqs. (13) and (16) we obtain the well known dispersion relation for the dielectric constant $\varepsilon(\omega)$

$$\frac{c^2 q^2}{\omega^2} = \varepsilon(\omega) = \frac{(\omega_L^2 - \omega^2) \varepsilon_\infty}{\omega_T^2 - \omega^2}$$

For $\omega = 0$, the Lyddane-Sachs-Teller relation

$$\frac{\omega_L}{\omega_T} = \sqrt{\frac{\varepsilon_s}{\varepsilon_\infty}} \qquad (18)$$

one obtains. $\varepsilon_s$ is the permitivity in the static electric field, while $\varepsilon_\infty$ is the permitivity at high frequencies. From Eq. (13) one obtains two branches for the polariton frequencies

$$\omega_\mu^2 = \frac{1}{2}\left(\omega_L^2 + \frac{c^2 q^2}{\varepsilon_\infty}\right) \pm \frac{1}{2}\sqrt{\left(\omega_L^2 - \frac{c^2 q^2}{\varepsilon_\infty}\right)^2 + \frac{4\omega_L c q |F|^2}{\hbar^2 \sqrt{\varepsilon_\infty}}} \qquad (19)$$

By using Eq. (18) this expression may be written as a function of $\omega_T$.

## IV. PHONON POLARITONS IN TWO-DIMENSIONAL LATTICES

As is well known there are no transverse vibrations (TO) in one-dimensional lattices. It is the simplest case that TO mode and the phonon-polaritons exist in the two-dimensional lattices[9]. The electromagnetic field only interacts with the vibration modes describing the relative displacement of the atoms in the basis and does not interact with that describing the mass center motion of the atom. It is assumed thgat the electromagnetic field propagates parallelly to **y** = 0 plane along the **z** – axis.. Only the s-polarization (**x** – direction) component of the electromagnetic field can couple with TO phonon mode of the two-dimensional lattice. The dispersion relation and the electric field strength of the phonon polaritons in two-dimensional diatomic ionic or polar crystal cannot be obtained by the conventional macroscopic dielectric theory of polaritons. A quantum mechanical treatment was presented in [9]. In this paper we apply the quantum mechanical theory presented in [3] So, in the above expressions now N is the number of unit cells per unit area, and in a square lattice

$$S = \left|\frac{1}{\sqrt{N}} e^{-i q z_n}\right|^2 = 2[2 + \cos(q_x a) + \cos(q_z a)] \qquad (20)$$

For $q_x a \ll 1$, $q_z a \ll 1$, this relation becomes

$S = 8 - q^2 a^2$

Consider now a $CuO_2$ planar lattice which arises in cuprate superconductors[14,15]. This is a lattice with a basis of three atoms: one of them of *Cu*, positionned at (0,0), with the mass $m_1$, one of oxygen, of mass $m_2$, positionned at (1/2,0) and the other oxigen positionned at (0,1/2). The lattice constant is 2*a*. We assume that the atoms oscillate perpendicular to the lattice plane. Then arise three transverse branches of oscillations: one is transverse acoustic (TA) phonon, of frequency ω' and two are transverse optical (TO) phonons, of frequencies ω" and ω''', repectively. These frequencies are determined from the equation[16]

$$(\omega^2 - \omega_2^2)[\omega^4 - (\omega_2^2 - \omega_1^2)\omega^2 + \frac{\omega_1^2 \omega_2^2}{2}(\sin^2 q_x a + \sin^2 q_y a)] = 0 \qquad (21)$$

$$\omega_1^2 = \frac{4\beta}{m_1} \qquad \omega_2^2 = \frac{2\beta}{m_2}$$

where β is the force constant, which is of the order [15] β = 159 N/m, so that $\omega_1$ = 7.75×10$^{13}$ c/s, $\omega_2$ = 1.094×10$^{14}$ c/s. In the centre of the Brillouin zone $q_x = q_y = 0$, from Eq. (21) one obtains

$$\omega'=0(TA), \omega''=\omega_2(TO1), \omega'''=\sqrt{\omega_1^2+\omega_2^2}(TO2)=1.341\times10^{14} \text{ c/s}$$

and from Eq. (19) and (17), one obtains for polariton frequencies

$$\omega''_{p1}(TO1)=1.102\times10^{14} c/s, \omega''_{p2}(TO1)=0$$

$$\omega'''_{p1}(TO2)=1.348\times10^{14} c/s, \omega'''_{p2}(TO2)=0$$

At the point M(11) of the Brillouin zone, $q_x = q_y = \pi/2a$, one obtains

$$\omega'=\omega_1(TA), \omega''=\omega'''=\omega_2(TO1\wedge TO2)$$

and from Eq. (19) one obtains for polariton frequencies

$$\omega''_{p1}=\omega''_{p1}=1.102\times10^{14} c/s, \omega''_{p2}\omega'''_{p2}=2.55\times10^{18} c/s$$

At the point X(10), $q_x = \pi/2a$, $q_y = 0$, one obtains

$$\omega'=\omega_2(TA), 2\omega''^2=\omega_1^2+\omega_2^2+\sqrt{\omega_1^4+\omega_2^4}$$

$\omega''(TO1)$ = 1.252×10$^{14}$ c/s and $\omega_{p1}''$ = 1.257×10$^{14}$ c/s. $\omega_{p2}''$ = 1.8×10$^{18}$ c/s. Further,

$$2\omega'''^2=\omega_1^2+\omega_2^2-\sqrt{\omega_1^4+\omega_2^4}$$

$\omega'''(TO2)$ = 4.789×10$^{13}$ c/s and $\omega_{p1}'''$ = 4.99×10$^{13}$ c/s, $\omega_{p2}'''$ = 1.8×10$^{18}$ c/s. The TO1 branch corresponds to the vibrations of the two oxygen atoms in opposite phase. We have evaluated $\omega_L$ from Eq. (17) by assuming that $\varepsilon_\infty$ = 1. It is obtained that $\omega_L$ is by one percent larger than $\omega_T$.

**V. PLASMONS**

We now consider briefly the dielectric constant associated with the uniform plasma mode[10]. In this case we replace $\omega_L = \omega_p$ (resonance frequency of plasmons) and $\omega_T = 0$ in Eq. (14) and we take

$$\omega_p^2 - \frac{4\omega_p|F|^2\sqrt{\varepsilon_\infty}}{\hbar^2 cq}=0 \qquad (22)$$

Therefore, Eq. (13a) becomes

$$\omega^2 - \omega_p^2 - \frac{c^2q^2}{\varepsilon_\infty}=0$$

and

$$\varepsilon(\omega) = \frac{c^2 q^2}{\omega^2} = \varepsilon_\infty - \frac{\omega_p^2}{\omega^2} \varepsilon_\infty \qquad (22a)$$

In the expression (101) of F we replase $\rho = \hbar q/2Rc$, $\omega_k = ck$, $\omega_{k+q} = c(k+q)$, $\Omega_q = \omega_p$, $k = q$, where $m$ is the electron effective mass. In this case Eq. (22) becomes

$$\omega_p^2 = \frac{8 D^2 R^2 \sqrt{\varepsilon_\infty}}{c \hbar m q} S \quad ; \quad S = \left| \frac{1}{\sqrt{N}} \sum_n e^{-i q z_n} \right|^2$$

This is the situation when a plasmon absorbs and emits a photon. If in Eq. (10a) we replace $\rho = m/2R$, $\omega_k = \omega_{k+q} = \omega_p$, $\Omega_q = cq$, $m = \hbar q/c$, then Eq. (22) becomes

$$\omega_p^3 = \frac{16 D^2 R^2 q \sqrt{\varepsilon_\infty}}{m^2 c} S$$

This is the situation when a photon absorbs and emits a plasmon. The two processes are competitive, so that from these two last equations one obtains

$$\omega_p^{5/2} = \frac{8\sqrt{2} D^2 R^2 \sqrt{\varepsilon_\infty}}{c \hbar^{1/2} m^{3/2}} S \qquad (23)$$

We have used[3] Eq. (23) to determine the plasmon-photon coupling constant, D in some materials. The standard expression for the plasma frequency is that obtained in the Drude model

$$\omega_p = \sqrt{\frac{4 \pi n e^2}{m}} \qquad (24)$$

where $n$ is the electron density. The dispersion relation for plasmons is

$$\omega^2 = \omega_p^2 + \frac{8}{5} \frac{E_F q^2}{m} \qquad (24a)$$

where $E_F$ is the Fermi energy level. The last term from this expression is of the order of $10^{-4} \omega_p^2$ for the values of $q$ enclosed in the first Brillouin zone. This expression, as well as the expression (23), is obtained by using the equation of the interaction of the plasmon with photons. Now, when plasmons are excited by electrons, may occur the following two situations: When an electron absorbs and emits a plasmon, then we replace in Eq. (10a) $\rho = m/2R$, $\omega_k = \omega_{k+q} = \omega_p$, $\Omega_q = \hbar q^2/2m$, $k = q$, and Eq. (22) becomes

$$\omega_p^3 = \frac{16 D^2 R^2 q \sqrt{\varepsilon_\infty}}{m^2 c} S \qquad (25a)$$

When a plasmon absorbs and emits an electron, then we replace in Eq. (22) $\rho = m/2R$, $\omega_k = \hbar k^2/2m$, $\omega_{k+q} = \hbar(k+q)^2/2m$, $\Omega_q = \omega_p$, $k = q$, and E#q. (22) becomes

$$\omega_p^3 = \frac{8 D^2 R^2 \sqrt{\varepsilon_\infty}}{\hbar mcq} S \qquad (25\,b)$$

The two processes are competitive, so that from eQs (25a) and (25b) one obtains, the same equation (23). The expression (25) is valid when $(q,\omega)$ is inside the band which is produced by the spectrum of the electrons. The scattering between plasmons and electrons is possible and will lead to damping of plasmons. It is possible that experimentally may be evidenced the plasma frequency given by Eq. (25a), if is dominant the process where one electron absorbs and emits a plasmon.

Now we apply Eq. (23) to graphene. The nearest neighbor vectors on the hexagonal lattice are

$$\boldsymbol{a}_1 = a_o\left(\frac{\sqrt{3}}{6}, 1\right) \; ; \qquad \boldsymbol{a}_2 = a_o\left(\frac{\sqrt{3}}{6}, -1\right) \; ; \qquad \boldsymbol{a}_3 = \left(\frac{-\sqrt{3}}{3}, 0\right)$$

and

$$S = 2\left[3 + 2\cos\left(\frac{\sqrt{3}}{6} a_o q_x\right) \cos(a_o q_y) + \cos\left(\frac{\sqrt{3}}{3} a_o q_x\right)\right] \qquad (26)$$

In a two dimensional metal with a uniform distribution of the electrons, we have

$$S = \frac{Z_n}{A} \int_0^R r\, dr \int_0^\pi e^{iqr\cos\theta} \sin\theta\, d\theta = \frac{Z_n}{qR} J_1(qR) \qquad (27)$$

where $Z_n$ is the number of electrons enclosed in a circle of area $A = \pi R^2$ and $J_1$ is the Besssel function of first kind. $R = r_s$ is the distance between two electrons.
In a three dimensional metal

$$S = \frac{Z_n}{V} 2\pi \int_0^R r^2\, dr \int_0^\pi e^{iqr\cos\theta} \sin\theta\, d\theta = \frac{3 Z_n}{(qR)^3}\left[\sin(qR) - qR\cos(qR)\right] \qquad (28)$$

where

$$R = r_s = \left(\frac{3}{4\pi} \frac{V}{N}\right)^{1/3}$$

is the length scale characterizing the distance between electrons. $N$ is the number of electrons enclosed in volume $V$. In Table I we present the calculated values of the plasmon frequencies, $\omega_p$, for graphene, Cu(2D) and Cu(3D) respectively, by using formulae (23) and (26)-(28). In Eq. (26) we have taken $q_y = 0$. We have used the following values of the constantes. In graphene, $D = 397.04$ N/m, $R = 1.42$ Å, in Cu(2D), $D = 291.2$ N/m, $R = 1.34$ Å, and in Cu (3D), $D = 291.2$ N/m, $R = 1.34$ Å. It is observed that due to the dependence of $S$ of $x = qR$, there is a slow decreasing of $\omega_p$ with $x$ increasing.

Table I. Plasmon frquency as a function of $x = qR$. In graphene, Cu(2D) and Cu(3D)

| $x$ | $\omega_p$(graphene) $\times 10^{16}$ c/s | $\omega_p$(Cu2D) $\times 10^{16}$ c/s | $\omega_p$(Cu3D) $\times 10^{16}$ c/s |
|---|---|---|---|
| 0 | 3.0407099 | 1.1325901 | 2.3191795 |
| 0.1 | 3.0405603 | 1.1325901 | 2.318246 |
| 0.2 | 3.0396964 | 1.128046 | 2.3154638 |
| 0.3 | 3.0384205 | 1.1265253 | 2.3108289 |
| 0.4 | 3.0366561 | 1.1234747 | 2.304349 |
| 0.5 | 3.0343262 | 1.1179513 | 2.299458 |
| 0.6 | 3.03159 | 1.114577 | 2.2858446 |
| 0.7 | 3.0282979 | 1.1051024 | 2.2738437 |
| 0.8 | 3.0248031 | 1.0966222 | 2.2600775 |
| 0.9 | 3.0201952 | 1.0869215 | 1.2443743 |
| 1.0 | 3.0153882 | 1.0761327 | 2.2269255 |
| 1.1 | 3.0100752 | 1.0644764 | 2.2076816 |
| 1.2 | 3.0042577 | 1.0512451 | 2.1866543 |
| 1.3 | 3.9979366 | 1.0374702 | 2.1866543 |
| 1.4 | 2.9911122 | 1.0224292 | 2.1392945 |

These values depend on the choice of the value of the coupling constant $D$.

## VI. SURFACE PLASMON POLARITONS

Surface plasmon polaritons (SPPs) are infrared or visible frequency electromagneic waves, which travel along a metal-dielectric or metal-air interface. SPPs can be excited by both electrons and photons. Excitation by electrons is generated by firing electrons into the bulk of the metal. As the electrons scatter, energy is transferred into the bulk plasma. The component of the scattering vector parallel to the surface results in the formation of a surface plasmon polariton[17,18]. In the case of excitation by photons, the coupling of photons into SPPs can be achieved using a coupling medium such as a prism or grating to match the photon and surface plasmon wave vector. By solving Maxwell equations for the electromagnetic waves at an interface between two materials with relative dielectric functions $\varepsilon_m$ (metal), $\varepsilon_d$ (dielectric) with the appropriate continuity relation the boundary conditions are

$$\frac{k_{zd}}{\varepsilon_d}+\frac{k_{zm}}{\varepsilon_m}=0 \qquad (29)$$

and

$$k_x^2+k_{zd}^2=\varepsilon_d\frac{\omega^2}{c^2} \qquad (30a)$$

$$k_x^2+k_{zm}^2=\varepsilon_m\frac{\omega^2}{c^2} \qquad (30b)$$

where at the interface metal-dielectric $k_{xm} = k_{xd}$. From the above equations one obtains the dispersion relation for surface polaritons

$$k_x^2 c^2 = \omega^2 \frac{\varepsilon_m \varepsilon_d}{\varepsilon_m+\varepsilon_d} \qquad (31)$$

By inserting relation (22a) into (31) one obtains the dispersion relation for surface plasmon polaritons

$$\omega^2 = \frac{1}{2\varepsilon_\infty \varepsilon_d}\{k_x^2 c^2(\varepsilon_\infty+\varepsilon_d)+\varepsilon_\infty \varepsilon_d \omega_p^2 - [(k_x^2 c^2(\varepsilon_\infty+\varepsilon_d)+\varepsilon_\infty \varepsilon_d \omega_p^2)^2 - 4\varepsilon_\infty^2 \varepsilon_d k_x^2 c^2 \omega_p^2]^{1/2}\} \qquad (32)$$

In Table II are presented the values of $\omega = f(x)$, in Cu (2D)/air and Cu(3D)/air interfaces as it resuits from formula (32). It is observed that $\omega$ presents at low values of $x$ a rapidly increasing, and then has a slow decreasing with $x$ increasing. The decrease follows the behavior of $\omega_p(x)$.

Table II. Frequency of surface plasmon polaritons as a function of $x = qR$, in Cu(2D)/air and Cu(3D)/air interfaces.

| X | $\omega$(Cu2D/air), ×10$^{16}$c/s | $\omega$(Cu3D/air) ×10$^{16}$c/s | x | $\omega$(Cu2D/air) ×10$^{16}$c/s | $\omega$(Cu3D/air) 10$^{16}$c/s |
|---|---|---|---|---|---|
| 0 | 0 | 0 | 0.8 | 0.7755643 | 1.5979675 |
| 0.1 | 0.8005779 | 1.637453 | 0.9 | 0.7685018 | 1.5717824 |
| 0.2 | 0.7975901 | 1.6366735 | 1.0 | 0.7609205 | 1.5682954 |
| 0.3 | 0.7965551 | 1.633738 | 1.1 | 0.7516648 | 1.5604485 |
| 0.4 | 0.7943551 | 1.6292636 | 1.2 | 0.7449832 | 1.5459625 |
| 0.5 | 0.7905694 | 1.6244845 | 1.3 | 0.7348469 | 1.5297059 |
| 0.6 | 0.78780355 | 1.616323 | 1.4 | 0.7245688 | 1.5132746 |
| 0.7 | 0.7813450 | 1.6070159 | | | |

## VII. CONCLUSIONS

We have continued to study the plasmons and polaritons by using the model presented in [3]. First, we attempt establish a correspondence between the vector potential **A** and the displacement **u** of an oscillator. The commutation relations for photons originates from the simple harmonic oscillator that are used to quantize the electromagnetic filed. Also, photons, phonons and plasmons are field quanta: photons are quanta of a continuous filed, while phonons and plasmons are quanta of a discrete field. So, we justify the using of Hamiltonian (10) for interaction between phonons and photons and plasmons and photons. We have studied the frequency of phonon polaritons in two dimensional lattice, with application to $CuO_2$ lattice. Further, we have evaluate plasma frequencies in graphene and in $Cu2D$ and $Cu3D$ metals. Finally, we evaluate the frequency of surface plasmon polaritons at the $Cu2D$/air and $Cu3D$/air interfaces. The obtained results are in a good agreement with experimental data. Our assertions are based on the fact that then when interact two systems, the result is determined by the system with more constraint rules.

## VII. APPENDIX. SIMILARITIES AND DIFFERENCES BETWEEN PHONONS, PLASMONS AND PHOTONS

Although photons, phonons and plasmons are all bosons, they do have significant defferences. For example phonons have associated the mass $M$ of an atom, plasmons have associated the mass $m$ of an electron, while the photons have zero rest mass. In order to be applied our model presented above, we have taken into account that photons have a "mass of motion", $\hbar q/c$.

Photons are called simple bosons because they are simple quanta with no internal structure. Phonons describe collective displacements of very many atoms, and plasmons describe plasma oscillations of the free electron gas density with respect to the fixed positive ions in a crystal. Phonons and plasmons are called composite bosons[19]. It is considered that there are two categories of bosons: type I referring to those bosons of an even number of fermions (such as $^4$He atoms), and type II referring to those that are collective excitations – such as phonons, plasmons, excitons, magnons, etc. However, because photons are the energy quanta of electromagnetic field modes, it is also possible to consider photons to be type II composite bosons.

Phonons, as well as plasmons, in general interact with each other. Photons interact very weak in vacuum and at low intensity. They interact with each other in nonlinear media ; their interaction is mediated by the atoms.

Phonons exist in discrete media, so that they have cut-off frequencies, which put an upper limit to their energy spectra. Photons do not have an upper limit for their energy. Plasmons have an upper limit for their energy.

Photons in free space have a linear dispersion relation. Phonons, as well as plasmons, have nonlinear dispersion relation.

Photons have two modes for each **k**. Phonons have 3 acoustic modes and (3$s$ – 3) optical modes for each **k**, where $s$ is the number of atoms per unit cell of the crystal. If in Eq. (24$a$) the $k$ term is negligible, then may be considered that plasmons have a uniform mode, whose resonance frequency is given by relatio0n (24).

Photons have the spin equals 1, while for phonons and plasmons spin is not defined. Some authors take the spin zero for phonons.

Macroscopic description of phonons and plasmons is performed with the aid of the wave equation for elastic continuum, while macroscopic description of photons is performed with the aid of Maxwell equations. However, in Section II we have established a correspondence between these two descriptions. In fact, their commutation relations originate from the simple harmonic oscillators that are used to quantize the electromagnetic field.

Phonons are quanta of a discrete field, while photons are quanta of a continuous field. Microscopic description of phonons and plasmons is performed with the aid of the Schroedinger equation, while the microscopic description of the photons is performed with the aid of quantum electrodynamics. However, we have shown[20,21] that in our model can be deduced the Coulomb's law by using a quantum mechanical treatment, with out taking into account of the electron charge. This result is out of the quantum electrodynamics.

REFERENCES

*Present address: University of Texas at Austin, Texas
[1]U. Fano, *Atomic theory of electromagnetic interactions in dense material*, Phts. Rev. **103,1**202-1218(1956)
[2]J. Hopfield, *Theory of contribution of excitonsto the complex dielectric constant of crystals*, Phys. Rv. **112**, 1555-1567(1958).
[3]V. Dolocan, A. Dolocan, V. O. Dolocan, *Some aspects of polaritons and plasmons in materials*, Commun Nonlinear Sci Numer Simulat **15**(1), 629-636 (2010)
[4]Ka-di Zhu, Zhong-long Xu, and Shi-wei Gu, *Quantum mechanical model calcuations of phonn polaritons in two-dimensional crystals*, Commun. Theor. Phys. (China Ocean Press), **14**, 385-395(1990).
[5]A. Hill, S. A. Mikhailov, and K. Ziegler, *Dielectric function and plasmons in graphene*, Euro. Phys. Letters, **87,** 27005(2009)
[6]J. M. Pitarke, V. M. Silkin, E. V. Chulkov, and P. M. Echenique, *Theory of surface plasmons and surface plasmon-polariton,*Rep. Prog. Phys. **70**, 1-87 (2007)
[7] L. C. Andreani, *Polaritons in multiple quantum wells*, Phys. Stat. Sol(b), **188**, 29-42(1995)
[8]A. J. Huber, B. Deutsch, L. Novotny, and L. Hillenbrabd, *Focusing of surface phonon polaritons*, Appl. Phys. Lettrs, **91,** 203104(2008)
[9]K. Leosson, *Optical Amplification of surface plasmon polaritons*, J. of Nanophotonics, **6**, 061801(2012)
[10]Stefan A.Maier, *Plasmons: Fundamentals and Applications*, Springer, 2007
[11]M. Balkanski, *Photon-phonon interactions in solids*, in Optical Properties of Solids, F. Abeles,ed., pp. 529-562, North Holland, Amsterdam, 1972
[12]J.C. Philips, *Ionicity of the chemical bond in crystals*, Rev. Mod. Phys., **42**, 312-356(1970)
[13]V. Dolocan, A. Dolocan and V.O. Dolocan, *Relation of interatomic forces in solids to bulk modulus, cohesive energy and thermal expansion*, Mod. Phys. Letters, **B22**, 2481-2492(2008)
[14]L. E. Dudy, *Nature and organization of the $CuO_2$ plane* , Dissertation, Humboldt-Universität zu Berlin, December 2008
[15]A. Punitha, Sujin P. Jose and S. Mohan, *Theoretical studies on phonon spectra of high temperature Tl-Ba-Ca-Cu-O superconductors*, J. of Physical Science,**22(2)**, 33-49(2011)
[16]J. Cazaux, *Initiation a la physique du solide*, Masson, 1996
[17]H. Lueth, *Solid Surfaces,Interfaces and Thin films,* Springer,2001
[18]S. Zeng,Yu. Xin, Law Wing-Keung, Zhang Yating, Hu Ruy, Dinh Xuan-Quyen, Ho. Ho-Puj, Yong Ken-Tie, *Size dependence of AuNP-enhanced surface plasmon resonance based on differential phase measurement,* Sensors and Actuators B: Chemical **176,** 1228(2012)

[19]W. Kohn, and D. Sherrington, *Two kinds of bosons and Bose condensates*, Rev. Mod. Phys. **42**, 1-39(1970)
[20]V. Dolocan, A. Dolocan and V. O. Dolocan, *Quantum mechanical treatment of the electron-boson interaction viewed as a coupling through flux lines*, International Journal of Modern Physics, **24**, 479-495(2010)
[21]V. Dolocan, *May be the electron taken off the charge?*, Romanian Reports in Physics, **64,** 15-23(2012)